\documentclass[preprint,showpacs,preprintnumbers,amsmath,amssymb,prb,aps]{revtex4-1}
\usepackage{epsfig}
\usepackage{epstopdf}
\usepackage{graphicx}
\usepackage{dcolumn}
\usepackage{bm}
\usepackage{slashbox}
\usepackage{textcomp}
\begin{document}

\title{Self-consistent evaluation of effective Coulomb interaction of V atom and its importance to understand the comparative electronic behaviour of vanadium spinels}
\author{Sohan Lal and Sudhir K. Pandey}
\affiliation{School of Engineering, Indian Institute of Technology Mandi, Kamand 175005, Himachal Pradesh, India}

\date{\today}

\maketitle

\section{Abstruct}

In present work, we try to understand the importance of effective Coulomb interaction ($U_{ef}$) between localized electrons of V atom to understand the comparative electronic behaviour of AV$_{2}$O$_{4}$ (A=Zn, Cd and Mg) compounds. The suitable values of $d$-linearization energy ($E_{d}$) of impurity V atom for calculating the $U_{ef}$ for these compounds are found to be $\geq$44.89 eV above the Fermi level. Corresponding to these values of $E_{d}$, the self-consistently calculated values of effective $U_{LSDA}$ ($U_{PBEsol}$) for ZnV$_{2}$O$_{4}$, MgV$_{2}$O$_{4}$ and CdV$_{2}$O$_{4}$ are $\sim$5.73 ($\sim$5.92), $\sim$6.06 ($\sim$6.22) and $\sim$5.59 ($\sim$5.71) eV, respectively. The calculated values of $\frac{t}{U_{ef}}$ ($t$ is the transfer integral between neighbouring sites) increases with decreasing V-V distance from CdV$_{2}$O$_{4}$ to MgV$_{2}$O$_{4}$ to ZnV$_{2}$O$_{4}$ and are found to be consistent with experimentally reported band gap. The values of $\frac{t}{U_{ef}}$ for ZnV$_{2}$O$_{4}$, MgV$_{2}$O$_{4}$ and CdV$_{2}$O$_{4}$ are found to be $\sim$0.023, $\sim$0.020 and $\sim$0.018, respectively. Hence, CdV$_{2}$O$_{4}$ with small (large) $\frac{t}{U_{ef}}$ (experimental band gap) as compared to ZnV$_{2}$O$_{4}$ and MgV$_{2}$O$_{4}$ is found to be in localized-electron regime, while ZnV$_{2}$O$_{4}$ and MgV$_{2}$O$_{4}$ are intermediate between localized and an itinerant-electron regime. The calculated values of lattice parameters $a_{LSDS}$ ($a_{PBEsol}$) are found to be $\sim$1.7\%, $\sim$2.0\% and $\sim$2.4\% ($\sim$0.6\%, $\sim$0.7\% and $\sim$0.7\%) smaller than $a_{exp}$ for CdV$_{2}$O$_{4}$, MgV$_{2}$O$_{4}$ and ZnV$_{2}$O$_{4}$, respectively, which indicates that the PBEsol functional predicts the lattice parameters in good agreement with the experimental data. The present study shows the importance of $U_{ef}$ in understanding the comparative electronic behaviour of these compounds.

\section{Introduction}

    In condensed matter physics, the local spin-density approximation (LSDA)/generalized gradient approximation (GGA) based on density functional theory has been one of the useful tool for understanding and predicting large range of properties of various materials\cite{Hohenberg,Kohn,Parr,Dreizler,Lundqvist,Mahan,Martin,Singh}. However, despite of the many successes, a clear limitation of both functionals have been seen in calculating the electronic and magnetic ground state properties of the strongly correlated systems\cite{Aryasetiawan}. For example, LSDA/GGA reproduces the ground state magnetic structure in the series NiO-MnO but fails to predict the insulating ground state of transition metal oxides such as CoO and FeO\cite{Terakura,Sawatzky}. Similarly, it fails to predict the anti-ferromagnetic insulating ground state as well as underestimate the magnitude of magnetic moment of La$_{2}$CuO$_{4}$ compound\cite{Pickett,Kasowski}. The drawbacks of LSDA/GGA method for describing the $d$ or $f$ electron systems are removed by adding Coulomb interaction ($U$). Hence, accurate LSDA+$U$/GGA+$U$ calculations for these materials depend on the selection of suitable value of $U$. Anisimov and Gunnarsson have discussed the meaning of parameter $U$ (earlier works given in their references), as the energy cost for moving a $d$-electron between two atoms having same number of electrons\cite{Anisimov}. $U$ corresponds to the parameter $F^{0}$ of unscreened Slater integrals in an atom. However, the effective $U$ in solids is much smaller than $F^{0}$ for atoms due to screening effect in the solids\cite{Anisimov}. Now in order to know the suitable value of effective $U$ for these systems, Anisimov and Gunnarsson have proposed a method based on the constrained density functional theory\cite{Anisimov}.   
     
    AV$_{2}$O$_{4}$ (A=Zn, Cd and Mg) compounds belong to the family of strongly correlated systems provide the large variety of interesting physical properties from last two decades\cite{Tsunetsugu,Tchernyshyov,Matteo,Motome,Maitra,Giovannetti,Pandey2011,Pandey2012,Lal2014,Lal2016,Lal126,Lee,Suzuki,Nishiguchi,Wheeler,Reehuis}. These spinels have the face-centered cubic structure at room temperature. A strong geometrical frustration arises in these compounds due to the corner-sharing network of octahedra of magnetically coupled V atoms\cite{Tchernyshyov,Tsunetsugu,Lee,Matteo,Wheeler,Nishiguchi,Mamiya,Reehuis,Suzuki,Chung}. Hence, these compounds remain in paramagnetic phase well below the Curie-Weiss temperature. The experimentally reported values of Curie-Weiss temperature for ZnV$_{2}$O$_{4}$ ($\sim$850 K)$>$MgV$_{2}$O$_{4}$ ($\sim$600 K)$>$CdV$_{2}$O$_{4}$ ($\sim$400 K) compounds\cite{Takagi}. Qualitatively, independent of non magnetic A site, these compounds show similar structural and magnetic behavior and undergo two phase transitions. First one is structural transition, which takes place from cubic to tetragonal phase of these compounds\cite{Mamiya,Nishiguchi,Reehuis,Onoda,Wheeler,Radaelli}. The experimentally observed  structural transition temperature for ZnV$_{2}$O$_{4}$, MgV$_{2}$O$_{4}$ and CdV$_{2}$O$_{4}$ are $\sim$50 K, $\sim$65 K and $\sim$97 K, respectively\cite{Nishiguchi,Reehuis,Wheeler,Takagi}. Second one is magnetic transition takes place from paramagnetic to anti-ferromagnetic phase\cite{Mamiya,Nishiguchi,Reehuis,Onoda,Wheeler,Radaelli}. The experimentally observed magnetic transition temperature for ZnV$_{2}$O$_{4}$, MgV$_{2}$O$_{4}$ and CdV$_{2}$O$_{4}$ are $\sim$40 K, $\sim$42 K and $\sim$35 K, respectively\cite{Nishiguchi,Reehuis,Wheeler,Takagi}. Based on localized-electron superexchange represented by $J$$\propto$$\frac{t^{2}}{U}$ (where, $t$ is the transfer integral between neighbouring sites), Canosa $et$ $al$. have shown that the magnetic transition temperature for these spinels increases with decrease in V-V separation. Hence they have assigned CdV$_{2}$O$_{4}$ to a localized-electron regime and ZnV$_{2}$O$_{4}$ and MgV$_{2}$O$_{4}$ to a intermediate between localized and an itinerant-electron regime\cite{Canosa}.    
    
    From above discussion, it is clear that the values of effective $U$ ($U_{ef}$) for these vanadates are expected to be different. Now, in order to understand the comparative electronic behaviour of these compounds, it is necessary to know the exact values of $U_{ef}$. Here, in present study we calculate the exact values of $U_{ef}$ for these spinels by using the first principles calculations. The values of $U_{ef}$ predicted by both LSDA and PBEsol functionals for MgV$_{2}$O$_{4}$$>$ZnV$_{2}$O$_{4}$$>$CdV$_{2}$O$_{4}$, when $d$-linearization energy of impurity V atom for these compounds is set to be $\geq$44.89 eV above the Fermi level. The calculated order of $\frac{t}{U_{ef}}$ for ZnV$_{2}$O$_{4}$$>$MgV$_{2}$O$_{4}$$>$CdV$_{2}$O$_{4}$ and is consistent with the experimentally reported order of band gap. Hence, CdV$_{2}$O$_{4}$ is set to be localized-electron regime, while ZnV$_{2}$O$_{4}$ and MgV$_{2}$O$_{4}$ are set to be intermediate between localized and an itinerant-electron regime. The calculated values of lattice parameter $a_{LSDS}$ ($a_{PBEsol}$) for CdV$_{2}$O$_{4}$, MgV$_{2}$O$_{4}$ and ZnV$_{2}$O$_{4}$ are $\sim$8.5436 ($\sim$8.6414), $\sim$8.2483 ($\sim$8.358) and $\sim$8.199 ($\sim$8.3435){\AA}, respectively and is consistent with the experimentally observed order of lattice parameter $a_{exp}$. 
   
\section{Computational details}

    In order to know the equilibrium values of lattice parameters and atomic coordinates of AV$_{2}$O$_{4}$ (A=Zn, Mg and Cd) compounds, we have performed the ferromagnetic calculations in the face centered cubic phase by using the full-potential linearized-augmented plane-wave (FP-LAPW) method as implemented in WIEN2k code\cite{Blaha}. The experimentally observed lattice parameters and atomic positions for these compounds are taken from the literature\cite{Onoda,Reehuis,Wheeler}. LSDA and GGA exchange-correlation functionals in the form of PW and PBEsol, respectively have been used in the present calculations\cite{Perdew1992,Perdew2008}. The muffin-tin sphere radii are set to 1.85, 2.1, 1.75, 2.0 and 1.6 Bohr for Zn, Cd, Mg, V and O atoms, respectively for every calculations. 8x8x8 k-point mesh size has been used here. Convergence target of total energy/cell and magnitude of force/cell for these systems have been set below 10$^{-4}$ Ry and 2 mRy/a.u., respectively. The equilibrium lattice parameters for these compounds are calculated by fitting the total energy difference between the volume dependent energies and energy corresponding to the equilibrium volume [$\Delta$$E$=$E$(V)-$E$(V$_{\rm eq}$)] per formula unit versus unit cell volume data using the universal equation of state\cite{Vinett}. The universal equation of state is defined as,
     
     $P$ = [3$B$$_{0}$(1 - $\chi$)/$\chi$$^{2}$]e$^{3/2(B'_{0}-1)(1-\chi)}$, $P$ = -($\partial$$E$/$\partial$$V$)   
    where $P$, $E$, $V$, $B$$_{0}$ and $B$$_{0}$$^{'}$ are the pressure, energy, volume, bulk modulus and pressure derivative of bulk modulus, respectively and $\chi$ = ($V$/$V_{0}$)$^{1/3}$. 
    
    In order to compute the numerical value of $U_{ef}$, Anisimov and Gunnarsson have constructed a general supercell approach with the hopping term (connecting the 3$d$ orbital of one atom with all other orbitals of remaining atoms) set to zero\cite{Anisimov}. $U_{ef}$ for correlated 3$d$ electrons are computed by varying the number of electrons in non-hybridizing 3$d$-shell by using following formula,\cite{Anisimov}
 
\begin{equation}    
    U_{ef}=\epsilon_{3d\uparrow}(\frac{n}{2}+\frac{1}{2},\frac{n}{2})-\epsilon_{3d\uparrow}(\frac{n}{2}+\frac{1}{2},\frac{n}{2}-1)-\epsilon_{F}(\frac{n}{2}+\frac{1}{2},\frac{n}{2})+\epsilon_{F}(\frac{n}{2}+\frac{1}{2},\frac{n}{2}-1)  
\end{equation}
    
where, $\epsilon$$_{3d\uparrow}$ and $\epsilon$$_{F}$ are the spin-up 3$d$ eigenvalue and the Fermi energy for the configuration of $n$ up-spins and $n$ down-spins, respectively. $n$ is the total number of 3$d$ electrons. Now, following the procedure given by Madsen and Nov$\acute{\rm a}$k, we have calculated the $U_{ef}$ of impurity V atom in these spinels by using WIEN2K code\cite{Madsen}. Theoretically computed and experimentally observed values of lattice parameters and atomic coordinates are used for calculating the $U_{ef}$. The values of $U_{ef}$ of impurity V atom in these spinels are calculated by using the Eqn. 1, where two calculations of 2x2x2 face centered supercell using 3x3x3 k-point mesh size are performed. In both calculations, one impurity V atom with the $d$-configuration forced to be as in Eqn. (1). The magnetic V ion has two 3$d$ electrons. Hence, one calculations with 1.5$\uparrow$, 1$\downarrow$ and second calculations with 1.5$\uparrow$, 0$\downarrow$ constrained $d$-electrons of impurity V atom (added to the core region) are performed. Also, the $d$ impurity electrons have been removed from the valence in both calculations. Now, in order to eliminate the $d$-character of the APWs at the impurity atom, we have varied the $d$-linearization energy ($E_{d}$) above the Fermi level. Convergence target of charge/cell for these systems has been set below 10$^{-2}$ electronic charge. The rest of the computational details are similar as described above.

\section{Results and discussions}

    ZnV$_{2}$O$_{4}$ and CdV$_{2}$O$_{4}$ compounds crystallize in face centered cubic spinel structure with the space group {\it Fd$\bar{3}$m}. In both compounds, the Wyckoff positions of (Zn, Cd) and V atoms are 8$a$ (0.125,0.125,0.125) and 16$d$ (0.5,0.5,0.5), respectively. The O atom is located at the Wyckoff position 32$e$ (x,x,x), where the experimentally observed values of x for both compounds are shown in the Table 1. However, MgV$_{2}$O$_{4}$ crystallizes in the face centered cubic structure having space group {\it F$\bar{4}$3m}. In this compound, the Wyckoff positions of Mg atom are 4$a$ (0,0,0) and 4$c$ (0.25,0.25,0.25). Both V and O (O1 and O2) atoms are located at the Wyckoff position 16$e$ (x,x,x), where the experimentally observed values of x are also shown in the Table 1.  
    
   In order to calculate the equilibrium values of lattice parameter $a$ for these compounds, we have plotted the total energy difference between the volume dependent energies and energy corresponding to the equilibrium volume [$\Delta$$E$=$E$(V)-$E$(V$_{\rm eq}$)] per formula unit with varying volume of the unit cell. The plot of $\Delta$$E$ versus volume for these compounds using LSDA and PBEsol functionals are shown in the Fig. 1(a-f). It is clear from the figure that both functionals give the parabolic behaviour for all these compounds. The volume corresponding to minimum energy gives the exact equilibrium volume for these compounds. Here, we have fitted the energy-volume data by using the equation of states formula of Vinett $et$ $al.$, which gives the exact values of equilibrium volume for these spinels\cite{Vinett}. The values of equilibrium volume predicted by LSDA for CdV$_{2}$O$_{4}$, MgV$_{2}$O$_{4}$ and ZnV$_{2}$O$_{4}$ are $\sim$4208, $\sim$3787 and $\sim$3719 bohr$^{3}$, respectively. Similarly, PBEsol gives the equilibrium volume $\sim$4354, $\sim$3940 and $\sim$3919 bohr$^{3}$ for CdV$_{2}$O$_{4}$, MgV$_{2}$O$_{4}$ and ZnV$_{2}$O$_{4}$, respectively. It is also clear from the figure that the equilibrium volume predicted by PBEsol is $\sim$3.4\%, $\sim$4.1\% and $\sim$5.4\% larger than LSDA for CdV$_{2}$O$_{4}$, MgV$_{2}$O$_{4}$ and ZnV$_{2}$O$_{4}$, respectively. The calculated values of equilibrium lattice parameters $a_{LSDS}$ and $a_{PBEsol}$ corresponding to the equilibrium volume along with the experimentally observed lattice parameter $a_{exp}$ for these compounds are shown in the Table 1. The values of $a_{LSDS}$ for CdV$_{2}$O$_{4}$, MgV$_{2}$O$_{4}$ and ZnV$_{2}$O$_{4}$ are 8.5436, 8.2483 and 8.199 {\AA}, respectively. Similarly, the values of $a_{PBEsol}$ are 8.6414, 8.358 and 8.3435{\AA} for CdV$_{2}$O$_{4}$, MgV$_{2}$O$_{4}$ and ZnV$_{2}$O$_{4}$, respectively. It is evident from the table that among these spinels, the calculated values of $a_{LSDS}$ and $a_{PBEsol}$ are largest for CdV$_{2}$O$_{4}$ and smallest for ZnV$_{2}$O$_{4}$ and are consistent with the order of $a_{exp}$ for these compounds. Also, the values of $a_{LSDS}$ ($a_{PBEsol}$) are $\sim$1.7\%, $\sim$2.0\% and $\sim$2.4\% ($\sim$0.6\%, $\sim$0.7\% and $\sim$0.7\%) smaller than $a_{exp}$ for CdV$_{2}$O$_{4}$, MgV$_{2}$O$_{4}$ and ZnV$_{2}$O$_{4}$, respectively. Such an underestimation of lattice parameter is expected here because both functionals in general underestimate the lattice parameters. Above discussion clearly shows that the values of $a_{PBEsol}$ are in good agreement with the values of $a_{exp}$ as compared to the $a_{LSDS}$ for these compounds. 
    
    The calculated values of atomic coordinates for O and V (only for MgV$_{2}$O$_{4}$) corresponding to the equilibrium values of lattice parameters $a_{LSDS}$ and $a_{PBEsol}$ are also given in the Table 1. It is clear from the table that the calculated values of x$_{LSDA}$ and x$_{PBEsol}$ (represent the x, y and z coordinates of the O atom) are deviated $\sim$0.6\% and $\sim$0.2\%, respectively from the x$_{exp}$ for ZnV$_{2}$O$_{4}$ compound. Similarly, x$_{LSDA}$ and x$_{PBEsol}$ of O atom for ZnV$_{2}$O$_{4}$ are deviated from the x$_{exp}$ by only $\sim$0.04\%. However, for MgV$_{2}$O$_{4}$, x$_{LSDA}$ (x$_{PBEsol}$) of O1 and O2 atoms are deviated $\sim$0.5\% ($\sim$0.4\%) and $\sim$0.06\% ($\sim$0.1\%), respectively from x$_{exp}$. Similarly, the deviation of x$_{LSDA}$ and x$_{PBEsol}$ for V atom is $\sim$0.02\% from the x$_{exp}$ for this compound.  
   
    Now, we calculate the $U_{ef}$ of impurity V atom for these spinels by using the Eqn. 1. The calculated values of $U_{ef}$ by using LSDA and PBEsol functionals (corresponding to theoretically computed values of lattice parameters and atomic coordinates) for all three compounds are shown in the Table 2. The calculated values of effective $U_{LSDA}$ ($U_{PBEsol}$) for ZnV$_{2}$O$_{4}$, MgV$_{2}$O$_{4}$ and CdV$_{2}$O$_{4}$ are $\sim$1.50 ($\sim$8.99), $\sim$9.65 ($\sim$9.33) and $\sim$8.41 ($\sim$8.24) eV, respectively when $E_{d}$ is $\sim$31.29 eV above the Fermi level. However, further increase in the $E_{d}$ shows the sharp decrease in effective $U_{LSDA}$ (except for ZnV$_{2}$O$_{4}$) and $U_{PBEsol}$ for these compounds. The values of effective $U_{LSDA}$ ($U_{PBEsol}$) are $\sim$5.73 ($\sim$5.92), $\sim$6.06 ($\sim$6.22) and $\sim$5.59 ($\sim$5.71) eV for ZnV$_{2}$O$_{4}$, MgV$_{2}$O$_{4}$ and CdV$_{2}$O$_{4}$, respectively as the $E_{d}$ is $\sim$44.89 eV above Fermi level. However, a small change in the values of effective $U_{LSDA}$ and $U_{PBEsol}$ are observed, when $E_{d}$ increases from 44.89 to 58.50 eV above the Fermi level for these compounds. Corresponding to this value of $E_{d}$, the effective $U_{LSDA}$ ($U_{PBEsol}$) for ZnV$_{2}$O$_{4}$, MgV$_{2}$O$_{4}$ and CdV$_{2}$O$_{4}$ are $\sim$5.78 ($\sim$5.92), $\sim$6.08 ($\sim$6.24) and $\sim$5.72 ($\sim$5.73) eV, respectively. Almost, a similar behaviour of $U_{ef}$ (small changes in values) is observed by using the experimentally observed values of lattice parameters and atomic coordinates for both LSDA and PBEsol functionals. Above discussion clearly shows that the inconsistent values of $U_{ef}$ for both functionals are observed when $E_{d}$ of impurity V atom is $\sim$31.29 eV above the Fermi level. Hence, it is important to know the exact values of $E_{d}$ of impurity V atom for calculating the $U_{ef}$.  
  
    In order to know the suitable values of $E_{d}$ of impurity V atom for these compounds, we have plotted the density of states (DOS) for $d$ states of impurity V atom. The plot of DOS for $d$ states of impurity V atom corresponding to 1.5$\uparrow$, 1$\downarrow$ $d$ electrons and 1.5$\uparrow$, 0$\downarrow$ $d$ electrons using LSDA functional are shown in Fig. 2. It is clear from the figure that the contribution of $d$ states to the DOS are finite around Fermi level, when $E_{d}$ is $\sim$31.29 eV above the Fermi level for both calculations of these compounds. Hence, the $d$-character of the APWs at the impurity V atom are not completely removed. However, the contribution of $d$ states to the DOS are not observed around Fermi level, when $E_{d}$ is $\sim$44.89 and 58.50 eV above the Fermi level for both calculations of these compounds. Hence, indicate the absence of $d$-character of the APWs at the impurity atom. Almost, a similar behavior is also observed by using PBEsol functional, when both experimental and calculated structural parameters are used. Above discussion clearly shows that the large or small values of effective $U_{LSDA}$ of impurity V atom for these compounds corresponding to $E_{d}$ which is $\sim$31.29 eV as compared to $\sim$44.89 and $\sim$58.50 eV above the Fermi level are due to the presence of $d$-character of the APWs at the impurity atom. Hence, the suitable values of $E_{d}$ of impurity V atom in these compounds are found to be larger than $\sim$31.29 eV above the Fermi level. Similarly, one can find the suitable values of $E_{d}$ or $E_{f}$ for any transition metal compounds by looking the DOS of $d$ or $f$ states of impurity sites. From above discussion, it is also clear that the exact calculated values of effective $U_{LSDA}$ and $U_{PBEsol}$ for MgV$_{2}$O$_{4}$$>$ZnV$_{2}$O$_{4}$$>$CdV$_{2}$O$_{4}$. The different values of effective $U$ for these compounds are due to the different V-V, V-O distance and V-O-V angles of VO$_{6}$ octahedra.   
    
    Now, in order to know the electronic behaviour of these compounds, we have calculated the ratio between nearest neighbour hopping integral ($t$) and $U$. The values of $t$ for these spinels are calculated by fitting the following equation,\cite{Saul}
\begin{equation}    
    J\approx-\frac{4t^{2}}{U}
\end{equation}   
   
where, $J$ is the nearest neighbour exchange coupling constant. The linear fit of $J$ versus 1/$U$ for these compounds are shown in the Fig. 3, where the data are taken from our earlier publication\cite{Lal126}. After fitting, the values of $t$ for ZnV$_{2}$O$_{4}$, MgV$_{2}$O$_{4}$ and CdV$_{2}$O$_{4}$ are found to be $\sim$137, $\sim$124 and $\sim$101 meV, respectively. Hence, among these spinels the smallest (largest) values of $t$ for CdV$_{2}$O$_{4}$ (ZnV$_{2}$O$_{4}$) are due to increase in the overlap integral (the donor and acceptor orbitals on neighboring atoms) with decreasing V-V distance from CdV$_{2}$O$_{4}$ to MgV$_{2}$O$_{4}$ to ZnV$_{2}$O$_{4}$ compound. The values of $\frac{t}{U_{ef}}$ for ZnV$_{2}$O$_{4}$, MgV$_{2}$O$_{4}$ and CdV$_{2}$O$_{4}$ are 0.023, 0.020 and 0.018, respectively, where the $U_{ef}$ (calculated using PBEsol) are taken from Table 2 when $E_{d}$ is $\sim$44.89 eV above the Fermi level. Similar behaviour of $\frac{t}{U_{ef}}$ for these spinels is also observed by using other values of $U_{ef}$. Here, it is interesting to compare our result with the result of Canosa $et$ $al.$ for these spinels. According to Canosa $et$ $al.$, the overlap integral in $t$ increases with decrease in V-V distance for these compounds. Hence, the order of $\frac{t}{U_{ef}}$ for MgV$_{2}$O$_{4}$$>$ZnV$_{2}$O$_{4}$$>$CdV$_{2}$O$_{4}$, where they have assigned CdV$_{2}$O$_{4}$ to a localized-electron regime and MgV$_{2}$O$_{4}$ and ZnV$_{2}$O$_{4}$ to a intermediate between localized and an itinerant-electron regime\cite{Canosa}. However, in the present study the order of $\frac{t}{U_{ef}}$ are found to be ZnV$_{2}$O$_{4}$$>$MgV$_{2}$O$_{4}$$>$CdV$_{2}$O$_{4}$. The ambiguity of $\frac{t}{U_{ef}}$ in the Canosa $et$ $al.$ and present work for MgV$_{2}$O$_{4}$ and ZnV$_{2}$O$_{4}$ is due to the following reasons. (1) Canosa $et$ $al.$ conclude that the values of $U_{ef}$ remains constant or decreases with decreasing V-V distance in these spinels\cite{Canosa}. However, in the present study the self-consistently calculated values of $U_{ef}$ shows the opposite behaviour. (2) V-V distance in ZnV$_{2}$O$_{4}$ is considered to be large as compared to MgV$_{2}$O$_{4}$ in the work of Canosa $et$ $al.$, which is opposite to experimental data used in the present work\cite{Canosa,Roger,Nishiguchi}. Above discussion clearly shows that the calculated values of $\frac{t}{U_{ef}}$ in present study decreases with increase in V-V distance for these compounds. It is important to note that experimentally observed values of band gap decreases as V-V distance decreases for these compounds\cite{Roger,Canosa}. Hence, the order of $\frac{t}{U_{ef}}$ in the present work are consistent with the order of experimentally observed band gap. Hence, CdV$_{2}$O$_{4}$ with small (large) $\frac{t}{U_{ef}}$ (experimental band gap) is allocated to a localized-electron regime as compared to MgV$_{2}$O$_{4}$ and ZnV$_{2}$O$_{4}$, where both of these compounds are found to be intermediate between localized and an itinerant-electron regime. Hence, present work shows the importance of $U_{ef}$ in understanding the comparative electronic behaviour of these compounds.

\section{Conclusions}

     The self-consistent evaluation of effective Coulomb interaction ($U_{ef}$) of V atom and its importance to understand  the comparative electronic behaviour of AV$_{2}$O$_{4}$ (A=Zn, Cd and Mg) compounds have been studied by using density functional theory. Among these compounds, the $U_{ef}$ (calculated by LSDA and PBEsol functionals) were found to be largest for MgV$_{2}$O$_{4}$ and smallest for CdV$_{2}$O$_{4}$, when $d$-linearization energy of impurity V atom was set to be $\geq$44.89 eV above the Fermi level. The values of $\frac{t}{U_{ef}}$ ($t$ is the transfer integral between neighbouring sites) and experimental band gap increase and decrease from CdV$_{2}$O$_{4}$ to MgV$_{2}$O$_{4}$ to ZnV$_{2}$O$_{4}$, respectively. Hence CdV$_{2}$O$_{4}$ was set to be localized-electron regime, while ZnV$_{2}$O$_{4}$ and MgV$_{2}$O$_{4}$ were set to be intermediate between localized and an itinerant-electron regime. The calculated values of lattice parameters and atomic coordinates using PBEsol functional were in good agreement with the experimental data as compared to LSDA functional for these spinels. At last we conclude that the present work shows the importance of $U_{ef}$ in understanding the comparative electronic behaviour of these compounds.        

\acknowledgments {S.L. is thankful to UGC, India, for financial support.}

\pagebreak

\begin{table}[ht]
\caption{Experimentally observed and theoretically computed values of equilibrium lattice parameters ({\AA}) and atomic coordinates by using both LSDA and PBEsol functionals for AV$_{2}$O$_{4}$ (A=Zn, Mg and Cd) compounds.}
\centering 
\begin{tabular}{p{2.8cm}p{2.4cm}p{2.4cm}p{2.4cm}p{6.8cm}}
\hline
\hline
{Compound}&{{\it a$_{exp}$}}&{{\it a$_{LSDA}$}}&{{\it a$_{PBEsol}$}}&{Atomic coordinates}\\[0.0ex]
&&\\[0.0ex]
\hline
ZnV$_2$O$_4$&8.4028&8.1990&8.3435& O: 32e(x,x,x) x$_{exp}$=0.2604,\\
&&&&  x$_{LSDA}$=0.2588 and x$_{PBEsol}$=0.2598 \\
CdV$_2$O$_4$&8.6910&8.5436&8.6414& O: 32e(x,x,x) x$_{exp}$=0.2672,\\
&&&& x$_{LSDA}$=0.2673 and x$_{PBEsol}$=0.2673\\
MgV$_2$O$_4$&8.42022&8.2483&8.3580& O1: 16e(x,x,x) x$_{exp}$=0.38623, \\
&&&& x$_{LSDA}$=0.3844 and x$_{PBEsol}$=0.3846\\ 
&&&&O2: 16e(x,x,x) x$_{exp}$=0.86623, \\
&&&& x$_{LSDA}$=0.8657 and x$_{PBEsol}$=0.8653 \\
&&&&V: 16e(x,x,x) x$_{exp}$=0.6251, \\
&&&&  x$_{LSDA}$=0.6250 and x$_{PBEsol}$=0.6250 \\[1ex]
\hline
\hline
\end{tabular}
\label{table:the exp}
\end{table}

\begin{table}[ht]
\caption{Effective Coulomb interaction parameter of impurity V atom predicted by LSDA and PBEsol (in bracket) functionals for various values of $d$-linearization energy ($E_{d}$) of impurity V atom above the Fermi level ($E_{F}$) for AV$_{2}$O$_{4}$ (A=Zn, Mg and Cd) compounds. These values are computed corresponding to the calculated values of structural parameters.}
\centering 
\begin{tabular}{p{3.0cm} p{4.4cm} p{4.4cm} p{4.4cm}}
\hline
\hline
{Compound}&{$E_{d}$}=({$E_{F}$}+31.29) eV&{$E_{d}$}=({$E_{F}$}+44.89) eV&{$E_{d}$}=({$E_{F}$}+58.50) eV\\[0.5ex]
&$U_{LSDA}$($U_{PBEsol}$) [eV]&$U_{LSDA}$($U_{PBEsol}$) [eV]&$U_{LSDA}$($U_{PBEsol}$) [eV]\\[0.5ex]
\hline
ZnV$_2$O$_4$&1.50 (8.99)&5.73 (5.92)&5.78 (5.92)\\
MgV$_2$O$_4$&9.65 (9.33)&6.06 (6.22)&6.08 (6.24)\\
CdV$_2$O$_4$&8.41 (8.24)&5.59 (5.71)&5.72 (5.73)\\ [1ex]
\hline
\hline
\end{tabular}
\label{table:the exp}
\end{table}

\begin{figure}
\caption{Total energy difference between volume dependent energies and energy corresponding to the equilibrium volume [$\Delta$$E$=$E$(V)-$E$(V$_{\rm eq}$)] per formula unit versus unit cell volume plots for (a) ZnV$_{2}$O$_{4}$ (LSDA), (b) ZnV$_{2}$O$_{4}$ (PBEsol), (c) MgV$_{2}$O$_{4}$ (LSDA), (d) MgV$_{2}$O$_{4}$ (PBEsol), (e) CdV$_{2}$O$_{4}$ (LSDA), and (f) CdV$_{2}$O$_{4}$ (PBEsol) compounds.}
\includegraphics{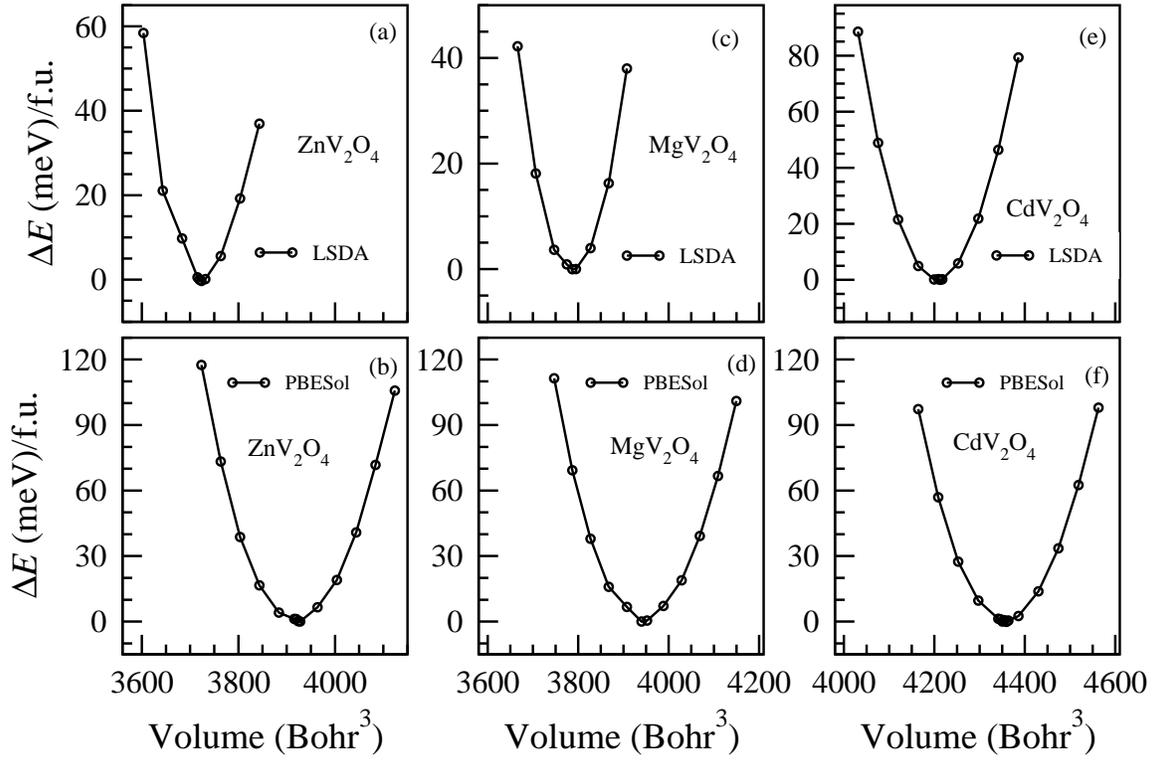}
\end{figure}

\begin{figure}
\caption{Density of states (DOS) for $d$ states of impurity V atom predicted by LSDA for (a) ZnV$_{2}$O$_{4}$ (1.5$\uparrow$, 1$\downarrow$ $d$ electrons), (b) ZnV$_{2}$O$_{4}$ (1.5$\uparrow$, 0$\downarrow$ $d$ electrons), (c) CdV$_{2}$O$_{4}$ (1.5$\uparrow$, 1$\downarrow$ $d$ electrons), (d) CdV$_{2}$O$_{4}$ (1.5$\uparrow$, 0$\downarrow$ $d$ electrons), (e) MgV$_{2}$O$_{4}$ (1.5$\uparrow$, 1$\downarrow$ $d$ electrons) and (f) MgV$_{2}$O$_{4}$ (1.5$\uparrow$, 0$\downarrow$ $d$ electrons) compounds. DOS is computed corresponding to the calculated values of structural parameters, where zero energy corresponds to the Fermi level.}
\includegraphics{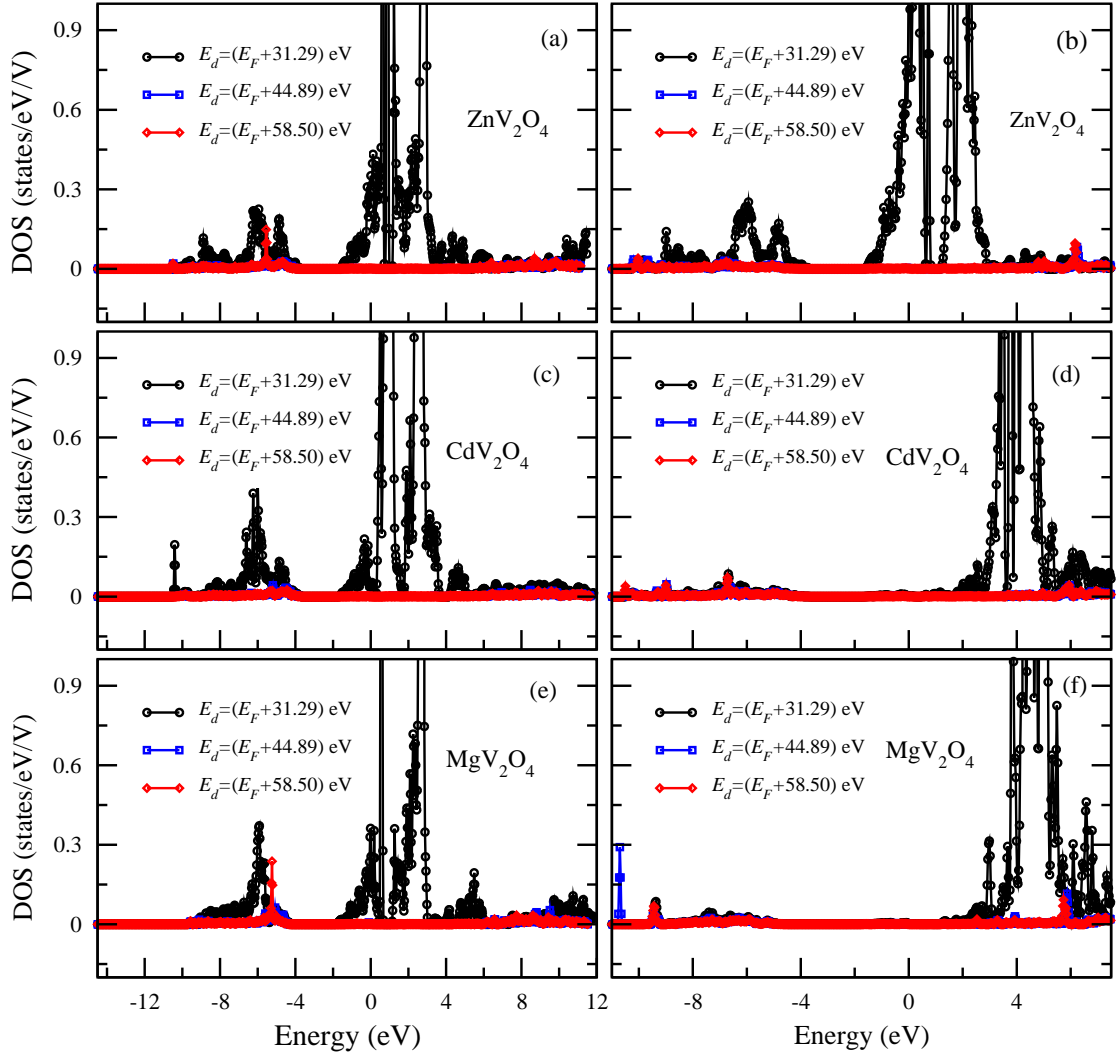}
\end{figure}

\begin{figure}
\caption{The nearest neighbour exchange-interaction parameter $J$ as a function of 1/$U$ for AV$_{2}$O$_{4}$ (A $\equiv$ Zn, Cd and Mg) compounds, where linear fit of $J$ versus 1/$U$ are shown by straight line.}
\includegraphics{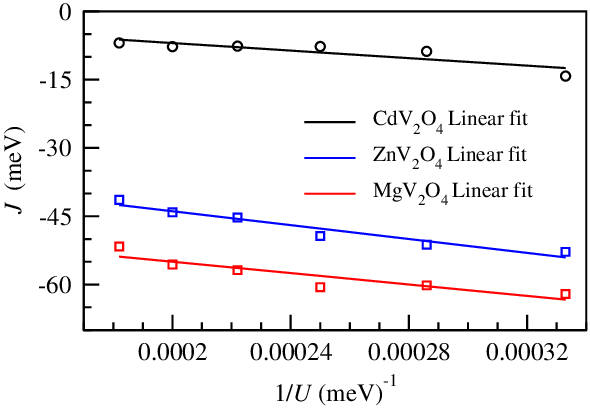}
\end{figure}

\end{document}